\documentclass[12pt]{iopart}

\usepackage{graphicx,setspace}
\usepackage{dsfont}
\usepackage{iopams}
\usepackage{mathrsfs}
\usepackage{psfrag}
\usepackage{cases}

\begin{document}

\title{All-frequency reflectionlessness}
\author{T G Philbin}
\address{Department of Physics and Astronomy, University of Exeter,
Stocker Road, Exeter EX4 4QL, UK.}
\eads{\mailto{t.g.philbin@exeter.ac.uk} }

\begin{abstract}
We derive planar permittivity profiles that do not reflect perpendicularly exiting radiation of any frequency. The materials obey the Kramers-Kronig relations and have no regions of gain. Reduction of the Casimir force by means of such materials is also discussed.
\end{abstract}


The propagation of electromagnetic waves in inhomogeneous materials remains a subject of intense interest and activity. This is partly because the challenge is not merely computational; there is also a need for improved analytical techniques. The issue of reflection by inhomogeneous materials illustrates the point well. It would be naively expected that inhomogeneity always causes some reflection and that this becomes very significant if the refractive index changes appreciably over a wavelength. But there are many inhomogeneous index profiles that have strictly zero reflection even when the geometrical-optics approximation is arbitrarily bad. The question of what governs reflection is subtle and a completely general answer is still elusive~\cite{ber90,hor15a}. 

Several infinite classes of reflectionless electromagnetic materials have been identified. The simplest of these classes from the theoretical viewpoint is provided by transformation optics, where {\it any} coordinate transformation gives a reflectionless inhomogeneous, anisotropic material with equal permittivity and permeability ($\varepsilon_{ij}(\boldsymbol{r})=\mu_{ij}(\boldsymbol{r})$)~\cite{dol61,pen06,geom}. The requirement for significant magnetic response and negligible absorption is a significant drawback of this class. There is however a reflectionless class of isotropic planar permittivity profiles $\varepsilon(x)$ in which absorption can be incorporated~\cite{hor15}. The requirement is for the function $\varepsilon(x)$, analytically continued to complex position values, to have no zeros or poles in the upper (or alternatively lower) half-plane~\cite{hor15}. The permittivity profile will then be reflectionless from one side, for all angles of incidence~\cite{hor15}. The criterion for this class also includes profiles $\varepsilon(x)$ with gain instead of loss, or profiles with regions of loss and gain. There are other reflectionless permittivity classes that necessarily feature both gain and loss for complex $\varepsilon(x)$, but also include purely real permittivity profiles~\cite{lek07,the14,lin11,reg12}. 

All of these classes are usually considered in the context of a monochromatic incident wave. If the reflectionless property is to be extended to a range of frequencies one must take account of dispersion and ensure that the material stays within the reflectionless class as the dielectric functions change with frequency. Particularly in the case of the class in~\cite{hor15}, where absorption is naturally included, one can envision engineering zero reflection over a significant frequency range using the recipe for $\varepsilon(x)$ with the parameters frequency dependent. The constraint imposed by Kramers-Kronig relations does not in principle restrict dispersion engineering over a finite frequency range, as the behaviour of $\varepsilon(x)$ outside the frequency range of interest can ensure that those relations are satisfied. For most applications only a limited range of frequencies is relevant, and this may be why there has been little consideration of the extent to which reflection at all frequencies can be eliminated. There is one application however where the reflection properties at all frequencies is the determining factor, namely the Casimir effect~\cite{lif55,dzy61,LLsp2}. Formulae for Casimir forces contain the reflection properties of the materials, usually as simple reflection coefficients, and these formulae involve an integration over all frequencies. In an important sense the Casimir effect is caused by reflection. The general question of reducing the Casimir force thus requires consideration of how much reflection can in principle be eliminated using artificial electromagnetic materials.

In considering reflection at all frequencies we must of course take full account of the significant constraint imposed by the Kramers-Kronig relations. For this reason a different approach is necessary compared to previous work on reflectionless materials. As noted at the outset, the properties that eliminate reflection are not fully understood, and here we take a rather blunt approach in order to derive some initial results on reflectionlessness at all frequencies.

We consider planar materials with no magnetic response (the latter offers no advantage since a  significant magnetic permeability is only realistic over narrow frequency ranges). The material is then described by a permittivity $\varepsilon(x,\omega)$ that is a function of one position coordinate and frequency. Physical considerations~\cite{LLcm} lead to the following constraints on the function $\varepsilon(x,\omega)$:
\begin{enumerate}
\item $\varepsilon(x,\omega)$ is analytic in the upper-half complex-$\omega$ plane (so that the  Kramers-Kronig relations hold).
\item  $\varepsilon(x,-\omega)=\varepsilon^*(x,\omega)$ (the susceptibility is real in the time domain).
\item The imaginary part of $\varepsilon(x,\omega)$ is positive for $\omega>0$ (no gain materials).
\item In the limit of zero frequency, $\varepsilon(x,\omega)\sim a(x)+b(x)\omega^n,\quad n\geq -1$.
\item In the limit of infinite frequency, $\varepsilon(x,\omega)\sim 1+c(x)/\omega^2$.
\end{enumerate}
We rule out gain materials by condition~(iii), as it is more interesting if effects can be achieved by passive materials. Conditions~(iv) and~(v)  come from the textbook account of the behaviour of dielectrics and metals at low and high frequencies~\cite{LLcm}. For metals at low frequencies the standard assumptions have been challenged, with claims that for Casimir calculations a plasma model permittivity ($\varepsilon\sim b/\omega^2$) rather than the Drude model ($\varepsilon\sim b/\omega$) should be used~\cite{kli09}. This issue turns out to be very important for eliminating reflection, as will be noted below. The detailed results reported here will be based on conditions~(i)--(v). 

It is presumably impossible, under conditions~(i)--(v), to obtain a permittivity $\varepsilon(x,\omega)$ that does not reflect waves from one direction for all frequencies and angles of incidence. The existence of such a permittivity would imply that the Casimir force can in principle be completely eliminated, since such a slab would experience no Casimir force from any materials positioned on the side from which it does not reflect. Next in order of interest would be a permittivity that does not reflect waves from one direction for all frequencies and for {\it one} angle of incidence. The natural choice in considering one angle is perpendicular incidence and we now restrict attention to this case.

For perpendicular incidence we do not need to distinguish between two independent polarizations of the wave and we can consider the scalar Helmholtz equation
\begin{equation}  \label{helm}
\left[\frac{d^2}{dx^2}+k_0^2\varepsilon(x,\omega)\right]E(x,\omega)=0, \qquad k_0=\frac{\omega}{c}.
\end{equation}
A simple method of generating materials that do not reflect from one side (utilized in~\cite{ber90,hor15a}, for example) is to write down an expression for $E(x,\omega)$ that for $x\to\pm\infty$ becomes a plane wave moving to the right (or left). One then substitutes $E(x,\omega)$ into (\ref{helm}) and solves for the material $\varepsilon(x,\omega)$. This method will usually fail for our purposes because the resulting $\varepsilon(x,\omega)$ will not satisfy conditions~(i)--(v). Nevertheless we pursue it, choosing an $E(x,\omega)$ with some degrees of freedom that we hope will allow us to meet the constraints on $\varepsilon(x,\omega)$.

Consider the field
\begin{equation}
\fl
E(x,\omega)=E_0 \exp\left[i k_0 \int_0^x dx'\, P(x',\omega) \right], \qquad P(x,\omega)\longrightarrow g_\pm(\omega) \quad \mathrm{as} \quad x \longrightarrow \pm \infty. \label{E}
\end{equation}
This is a right-going wave that propagates from one homogeneous region to another. Substitution of (\ref{E}) into (\ref{helm}) yields
\begin{equation} \label{ep}
\varepsilon(x,\omega)=\left[P(x,\omega)\right]^2-\frac{i c}{\omega}\frac{d}{dx} P(x,\omega).
\end{equation}
A choice of $P(x,\omega)$ for which (\ref{ep}) satisfies conditions~(i), (ii), (iv) and (v) is
\begin{equation}  \label{P}
P(x,\omega)=1+\frac{f(x)}{(\gamma-i\omega)^2}, \qquad \gamma>0,
\end{equation}
with real $f(x)$, as can be seen by inspection. As for condition~(iii), we note that with (\ref{P}) the imaginary part of (\ref{ep}) is
\begin{equation}
\fl
\mathrm{Im}\left[ \varepsilon(x,\omega) \right] = \frac{4 \gamma  \omega ^2 f(x) \left[\left(\gamma ^2+\omega ^2\right)^2+(\gamma^2 -\omega^2 ) f(x)\right]- c(\gamma^2 -\omega^2) \left(\gamma ^2+\omega ^2\right)^2 f'(x)}{\omega  \left(\gamma ^2+\omega ^2\right)^4}.  \label{imep}
\end{equation}
To see how to ensure this is positive for $\omega>0$, consider the limits $\omega\to 0$ and $\omega\to\infty$:
\begin{eqnarray}
\mathrm{Im}\left[ \varepsilon(x,\omega) \right] \to -\frac{c f'(x)}{\gamma^2\omega}, \qquad \omega\to 0,   \label{ep0}  \\
\mathrm{Im}\left[ \varepsilon(x,\omega) \right] \to\frac{4 \gamma  f(x)+cf'(x)}{\omega ^3}, \qquad \omega\to \infty.   \label{epinf}
\end{eqnarray}
The limit (\ref{ep0}) shows that $f(x)$ must monotonically decrease with $x$ ($f'(x)<0$), but then, from (\ref{epinf}), $4 \gamma  f(x)$ must exceed $|f'(x)|$ so in particular $f(x)>0$. As long as $\gamma$ is not too small these requirements arising from (\ref{ep0}) and (\ref{epinf}) are met by
\begin{equation}  \label{f}
f(x)=\frac{\Omega^2}{2}\left[1+\tanh(-x/a)\right], \qquad a>0.
\end{equation}
The choice (\ref{f}) also keeps $\mathrm{Im}\left[ \varepsilon(x,\omega) \right] $ positive throughout the entire range $\omega>0$, again if $\gamma$ is not too small. Checking this last fact is not too difficult but requires more careful inspection of (\ref{imep}). 

\begin{figure}[!htbp]
\begin{center} 
\includegraphics[width=10cm]{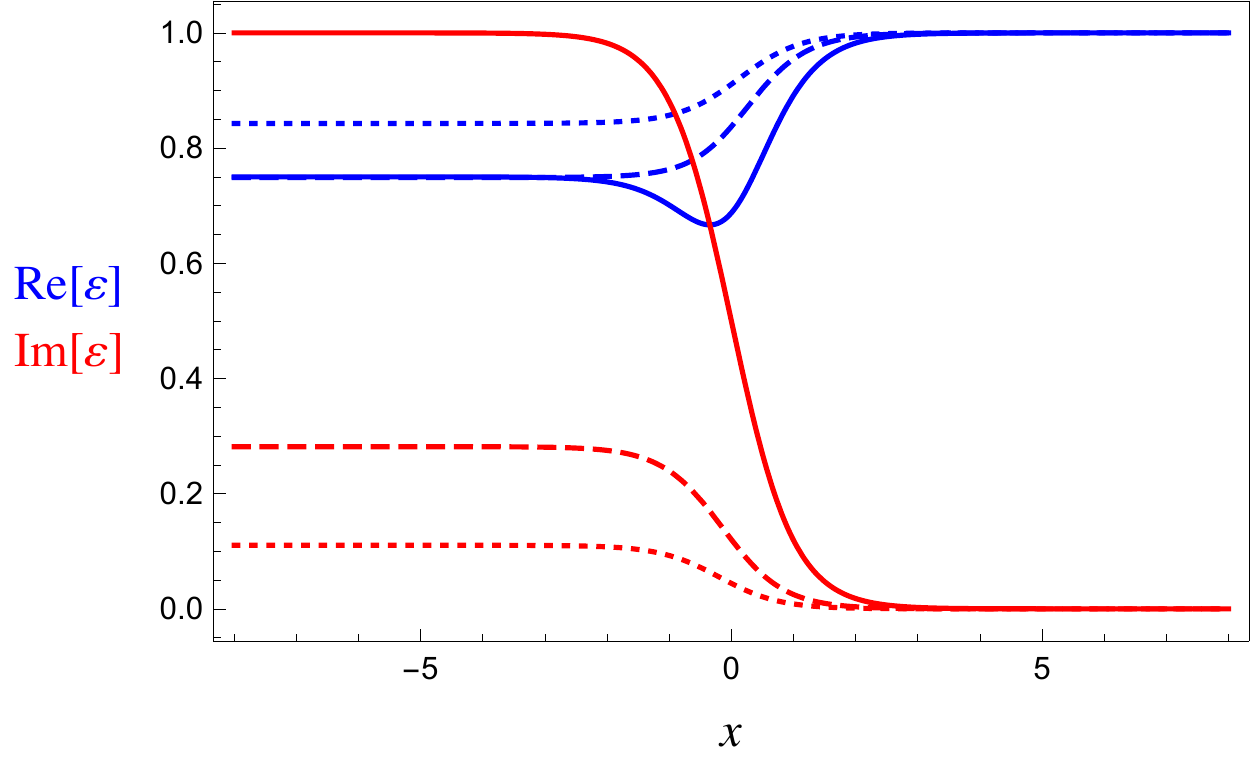}

\vspace{3mm}

\includegraphics[width=10cm]{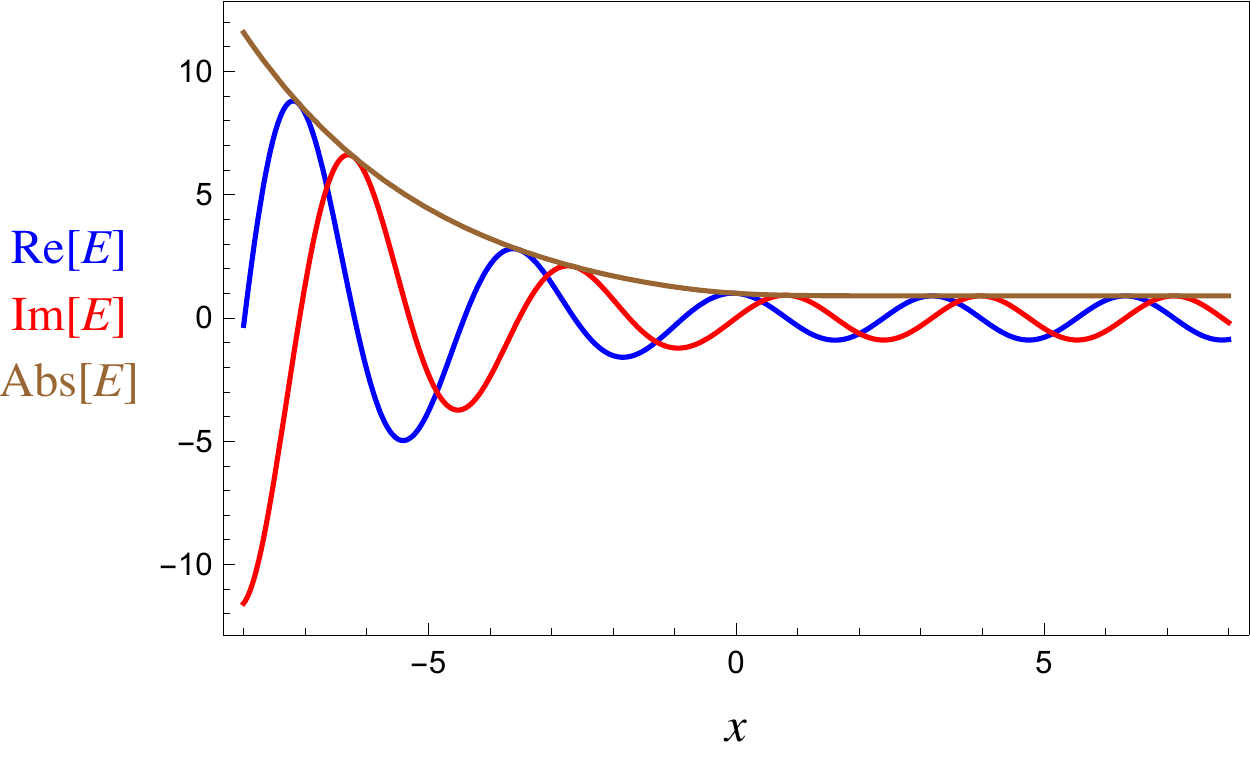}
\caption{The permittivity (\ref{ep}) and field (\ref{E}) as functions of position with $c=\gamma=\Omega=1$. Top: The real (blue) and imaginary (red) parts of the permittivity at three frequencies, $\omega=1$ (continuous lines), $\omega=2$ (dashed lines) and $\omega=3$ (dotted lines). Bottom: The real (blue), imaginary (red) and absolute value (brown) of the field of a right-travelling wave (\ref{E}) with frequency $\omega=2$ propagating in the permittivity profile. Any reflection of the wave would show up as ripples in $\mathrm{Abs}[E(x)]$. }
\label{fig:halfspace}
\end{center}
\end{figure}

With the above choices, the wave (\ref{E}) propagates from a homogeneous material region on the far left into vacuum on the far right, without any reflection (see Fig.~\ref{fig:halfspace}). Note we cannot make the boundary region in our example arbitrarily thin by taking $a\to0$ in (\ref{f}) as this would give a diverging permittivity. The requirement that $f(x)$ be monotonic has ruled out a solution with vacuum on both sides of the planar material. Also the solution is not for a wave incident on the material from outside, but rather a wave that exits the material into vacuum. The former case would be of more interest as we usually consider the reflection properties of waves incident on objects. Instead we have derived a planar material that does not reflect waves that exit it perpendicularly, for all frequencies. It might be thought that the solution above could be modified to derive a material that does not reflect waves perpendicularly incident from outside, but this is not possible. Changes such as time reversing the solution, which would work for that purpose, give a material with gain for all frequencies, violating condition~(iii). 

Alternatives to (\ref{E}) for a reflectionless wave in terms of some unknown function(s) were investigated but interestingly they all led to the same qualitative results. Thus under conditions~(i)--(v) it is not too difficult to find planar materials that do not reflect perpendicularly exiting waves of any frequency, but it is much more difficult to find materials that are reflectionless for perpendicularly entering waves of any frequency. Materials with the latter property are of course not ruled out by our very limited analysis. But it is not immediately apparent why the method chosen here should single out solutions of the ``exiting" type over those of the ``entering" type, if both types exist. 

We also mention that if condition~(iv) is changed to allow a $1/\omega^2$ divergence of the permittivity at zero frequency, then solutions of the ``entering" type exist and are not difficult to find using the method above. As the physical requirements for these solutions cannot be met using ordinary metals and dielectrics, we do not enter into details here (but see~\cite{kli09} on the possible relevance of the plasma model in the Casimir effect).

\begin{figure}[!htbp]
\begin{center} 
\includegraphics[width=10cm]{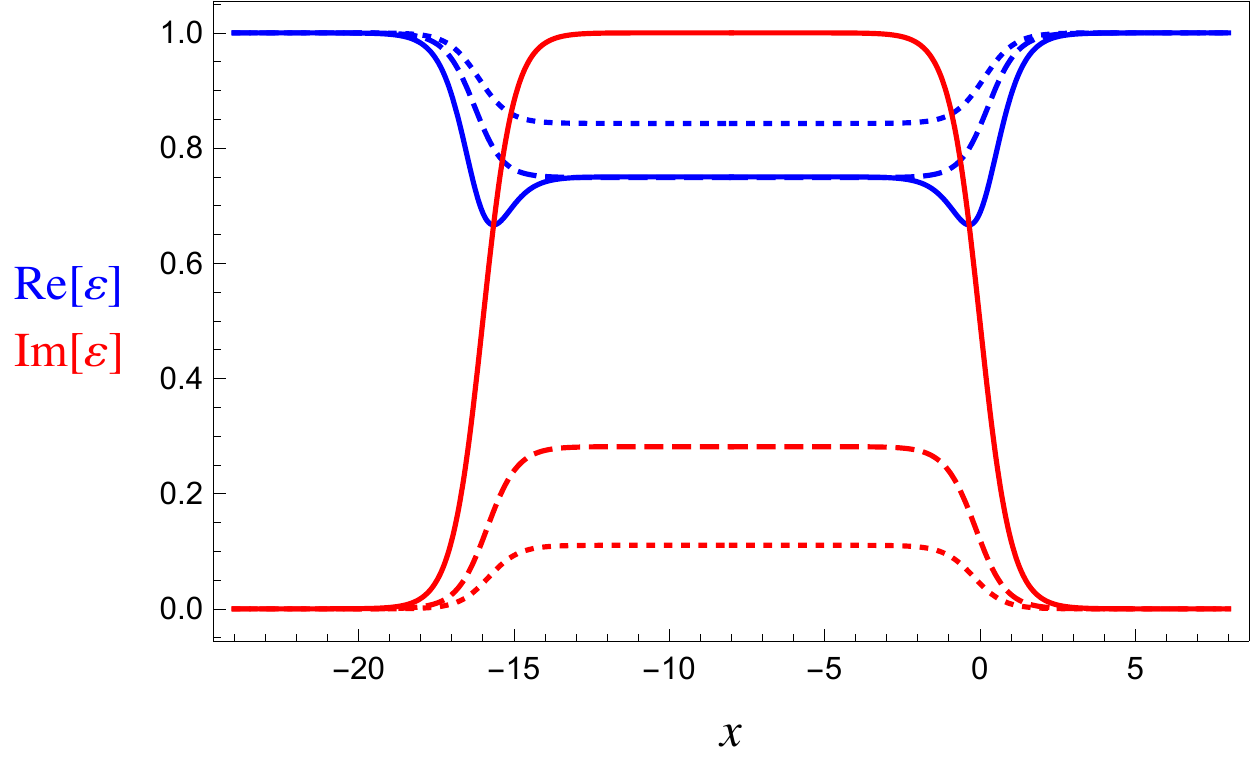}

\vspace{3mm}

\includegraphics[width=10cm]{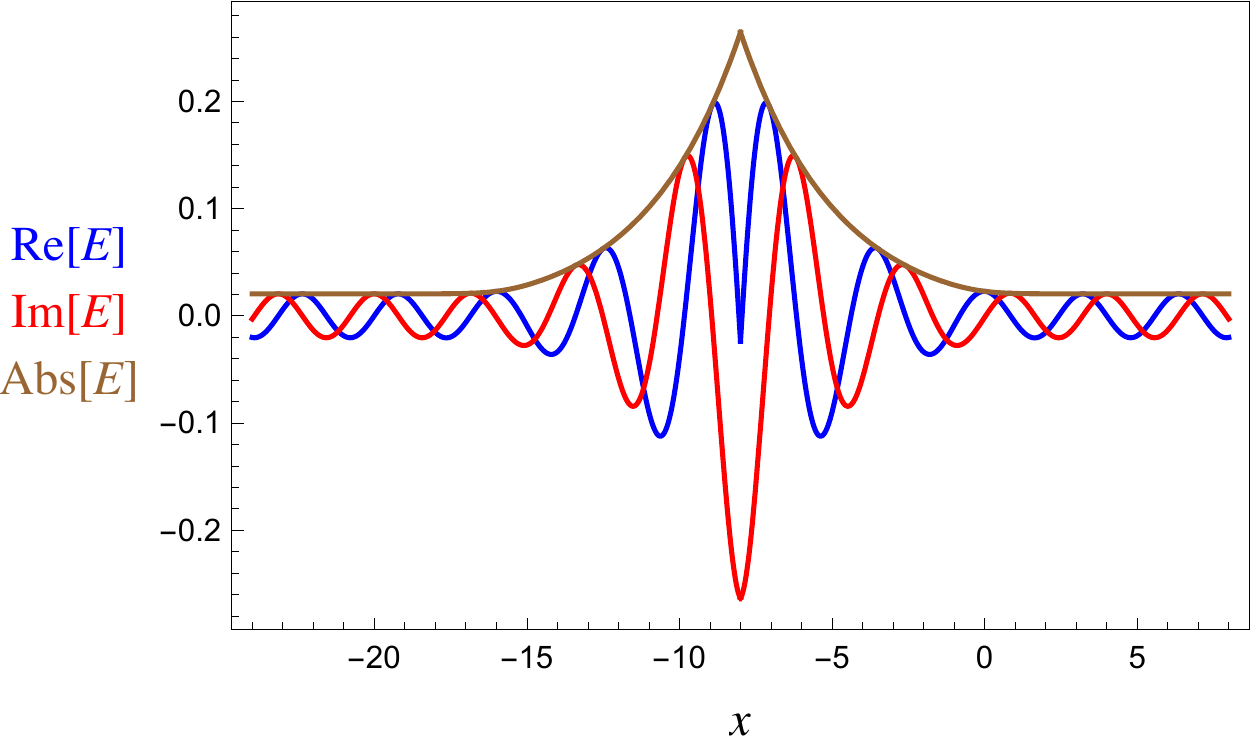}
\caption{Top: Permittivity of a planar slab formed by cutting the permittivity in Fig.~\ref{fig:halfspace} at $x=-8$ and joining to its mirror image in the plane of the cut. The plots correspond to those in Fig.~\ref{fig:halfspace}. Bottom:   Field of frequency $\omega=2$ produced by a plane source located at $x=-8$ in the slab. Waves propagate away from the source in both directions. There is no reflection of the waves as they exit into vacuum.  }
\label{fig:slab}
\end{center}
\end{figure}

It is easy to invert spatially the solution derived above so that the unreflected wave propagates to the left with homogeneous material on the far right and vacuum on the far left. We can also cut the material at some finite value of $x$ in the (approximately) homogeneous material region and attach its mirror image in the plane of the cut. The resulting material then approaches vacuum on each side and has the property that waves exiting perpendicularly through the inhomogeneous regions on either side are not reflected for any frequency (see Fig.~\ref{fig:slab}). The wave in Fig.~\ref{fig:slab} is that produced by a plane source in the centre of the slab; the amplitude and phase have been chosen so that in the homogeneous material region close to the source the wave matches the Green function $-i\exp(i\sqrt{\varepsilon} k_0 |x|)/(2\sqrt{\varepsilon} k_0)$ for equation (\ref{helm}) with constant permittivity. This material represents an ``inverse" of the usual parallel-plate arrangement considered in the Casimir effect: instead of matter-vacuum-matter we have vacuum-matter-vacuum. The latter geometry also leads to a Casimir force on the inhomogeneous regions forming the boundary of the planar slab (in the case of homogeneous slabs with sharp boundaries the forces act only on the boundary surfaces). The Casimir force in this case is determined by the reflection coefficients of waves exiting the slab on either side and it will be a compressing force pushing the boundary regions together~\cite{dzy61}. In our case however the reflection coefficients are zero for all frequencies for waves exiting perpendicular to the boundary regions, so there is no contribution to the Casimir force from the zero-point versions of these waves. As the angle of incidence changes from 90 degrees the reflection coefficients will increase continuously from zero and there will be a non-zero compressing Casimir force. By considering an effective 1D setup or waveguide it may be possible to perform a similar analysis to that given here and derive effective 1D electromagnetic materials that have no compressing Casimir force on their boundaries.

\ack
I thank S.\ Horsley for many useful discussions. Financial support was received from EPSRC under Program Grant EP/I034548/1.

\end{document}